\documentclass[preprint,12pt,3p]{elsarticle}

\usepackage{lineno,hyperref}
\modulolinenumbers[5]
\usepackage{booktabs} 
\usepackage{caption}
\usepackage{subcaption}
\usepackage[table,xcdraw]{xcolor}
\usepackage{tikz}
\usepackage{verbatim}
\def\checkmark{\tikz\fill[scale=0.4](0,.35) -- (.25,0) -- (1,.7) -- (.25,.15) -- cycle;} 
\journal{Journal of Information Processing and Management}

\begin{document}

\begin{frontmatter}

\title{Real Time Prediction of Drive by Download Attacks on Twitter}
	\author[label1]{Amir Javed\corref{cor1}}
		\address[label1]{School of Computer Science and Informatics, Cardiff University}
		
		\cortext[cor1]{I am corresponding author}
		
		\ead{javeda7@cardiff.ac.uk}
		
		\author[label1]{Pete Burnap}
		
		\author[label1]{Omer Rana}
		





\begin{abstract}
The popularity of Twitter for information discovery, coupled with the automatic shortening of URLs to save space, given the 140 character limit, provides cyber criminals with an opportunity to obfuscate the URL of a malicious Web page within a tweet. Once the URL is obfuscated the cyber criminal can lure a user to click on it with enticing text and images before carrying out a cyber attack using a malicious Web server. This is known as a {\it drive-by-download}. In a {\it drive-by-download} a user's computer system is infected while interacting with the malicious endpoint, often without them being made aware the attack has taken place. An attacker can gain control of the system by exploiting unpatched system vulnerabilities and this form of attack currently represents one of the most common methods employed. 
 In this paper we build a machine learning model using machine activity data and tweet meta data to  move beyond post-execution classification of such URLs as malicious, to predict a URL will be malicious with 99.2\% F-measure (using 10-fold cross validation) and 83.98\% (using an unseen test set) at 1 second into the interaction with the URL. Thus providing a basis from which to kill the connection to the server before an attack has completed and proactively blocking and preventing an attack, rather than reacting and repairing at a later date.
\end{abstract}

\begin{keyword}
\texttt Cyber security\sep Drive by Download\sep Malware\sep Machine Learning\sep Web Security

\end{keyword}

\end{frontmatter}


\section{Introduction}

Online social networks (OSNs) have emerged as powerful tools for disseminating information. Among these, Twitter, a micro-blogging website that allows its users to express themselves in 140 characters, has emerged as a go-to source for current affairs, entertainment news  and to seek information about global events in real-time. For example, Twitter has been used to study public reaction to events such as natural disasters \citep{Sakaki2009}, political elections \citep{Tumasjan2010} and terrorist attacks \citep{Burnap2014}. The England versus Iceland football match at the European Football Championships (Euro 2016) was one of the most tweeted about events of 2016 - attracting 2.1 million users \cite{Euro201657:online}. This high volume of users around a popular trending event and Twitter's inbuilt feature of shortening a URL due to its 140 character restriction provides cyber criminals with an opportunity to obfuscate links to malicious Web pages within \textit{tweets} and carry out a drive by download attack. In a {\it drive-by-download} \citep{Cova2010,moshchuk2006crawler} an attacker attempts to lure users to malicious Web pages so that they can hijack the user's system by exploiting a system vulnerability. By successfully carrying out these attacks an attacker is able to, for example, obtain remote access, steal user information, or make the computer part of a botnet \cite{provos2007ghost}.\\
The more popular OSNs become, the more attractive a platform they become for cyber criminals to conduct their attacks. Microsoft acknowledged this fast growing threat of malicious Web pages as one the top threats in their security and intelligence report published in 2013 \citep{microsoft2013} and the detection of drive by download attacks remains an important topic of research. The problem has been broadly investigated from a number of perspectives including: (i) characteristics of OSN user accounts (e.g. posting behaviours \citep{Cao2015} and social network links \citep{Yang2009}); (ii) characteristics of URLs (e.g. lexical features \citep{Ma2009a} and endpoint activity \citep{lee2013warningbird, Lee2011}); and (iii) analysing the code of a Web page in a static or dynamic manner to study its intended or actual behaviour when interacting with the underlying system on which the OSN user is accessing the Web page. 



The main contribution of this paper is a novel real time machine classification method that is based on the behavioural fingerprint of a shortened URL when an OSN user loads it into their web browser. By capturing machine activity metrics (e.g. CPU use, RAM use, Network I/O - for full list see Appendix 1) and tweet attributes, we are able to predict whether the URL is pointing to a malicious Web page with 99.2\% f-measure (using 10-fold cross validation) and 83.98\% (using an unseen test set) at 1 second into the interaction with a URL. This provides a novel contribution with which it is possible to kill the connection to the server before an attack has completed - thus proactively blocking and preventing an attack, rather than reacting and repairing at a later date. To the best of our knowledge this is the first study to proactively predict a drive by download attack by classifying a URL during interaction, rather than requiring the malicious payload to complete before classification. 

   
\section{Related work}

In this section we discuss the related work on the topic of detecting malicious content in online social networks. This is presented in two sub-sections - we first look at detecting such content using OSN user account and URL characteristics, and then study the use of static and dynamic code analysis. Using tweet meta-data Kristina and Rumi followed various top stories and used various tweet attributes to demonstrate how rapidly information (e.g. malicious URLs) can be disseminated in Twitter \citep{kristinalermanrumighosh2010} making it the core focus for existing work in this area - so the majority of the related work focusses on Twitter and tweet meta-data. It should be noted though that malicious URLs and spam are a significant issue on all OSNs. Table \ref{related} provides a summary of related work and the methods used at a high level for comparison.
\begin{table}[]
\centering
\caption{Malware or Spam detection techniques used}
\label{related}
\resizebox{\columnwidth}{!}{
\begin{tabular}{|l|l|l|l|l|l|l|}
\hline
\rowcolor[HTML]{C0C0C0} 
\multicolumn{7}{|c|}{\cellcolor[HTML]{C0C0C0}\textbf{Techniques used to detect Spam/Malware on Twitter}} \\ \hline
\rowcolor[HTML]{C0C0C0} 
\textbf{\begin{tabular}[c]{@{}l@{}}Methods Used \\ by Researchers\end{tabular}} & \textbf{Tweet Attributes} & \textbf{\begin{tabular}[c]{@{}l@{}}Blacklist \\ cross check\end{tabular}} & \textbf{\begin{tabular}[c]{@{}l@{}}Lexical analysis\\  of URL\end{tabular}} & \textbf{\begin{tabular}[c]{@{}l@{}}HoneyPot or \\ Honey profiles\end{tabular}} & \textbf{\begin{tabular}[c]{@{}l@{}}Machine Behaviour\\  (Network ,File,Process, \\ Memory,CPU etc)\end{tabular}} & \textbf{\begin{tabular}[c]{@{}l@{}}User Behaviour \\ on Twitter\end{tabular}} \\ \hline
\begin{tabular}[c]{@{}l@{}}OSN Account \\ Characteristics\citep{kristinalermanrumighosh2010} \citep{Cao2015} \citep{mohammadrezafaghanihosseinsaidi2009} \citep{chen2016}\\\citep{Stringhini2010} \citep{Benevenuto2010} \citep{Grier2010} 
 \citep{chaoyangrobertharkreaderjialongzhangseungwonshinguofeigu2012} \citep{Yang2009}\citep{Cresci2015}\end{tabular} & \checkmark{} & \checkmark{} &  & \checkmark{} &  & \checkmark{} \\ \hline
URL characteristics\citep{lee2013warningbird}\citep{Lee2011}\citep{Ma2009a} &  & \checkmark{} & \checkmark{} &  &  &  \\ \hline
\begin{tabular}[c]{@{}l@{}}Detect By analysing \\ Static Code\citep{McGrath2008}\citep{Canali2011} \citep{Kapravelos2013}\end{tabular} &  & \checkmark{} & \checkmark{} &  &  &  \\ \hline
\begin{tabular}[c]{@{}l@{}}Detect By analysing \\ Dynamic  Code\citep{Cova2010}\citep{kim2017j}\citep{jayasinghe2014efficient}\citep{wressnegger2016comprehensive}\\\citep{bartos2016optimized}\citep{Burnap2014}\citep{Cao2016}\end{tabular} & \checkmark{} &  &  &  & \checkmark{}(network only) &  \\ \hline
Our Model & \checkmark{} & \checkmark{} &  & \checkmark{} & \checkmark{} &  \\ \hline
\end{tabular}}
\end{table}

\subsection{Detecting Malicious Content based on OSN account and URL characteristics}

Previous research has aimed to identify tweets that are classified as spam or contain a URL pointing to a malicious Web server based on tweet meta-data. The rationale being that it is possible to differentiate between a 'normal' user and that of a cyber criminal based on user account characteristics extracted from meta-data such as number of followers, number of people they follow, their posting behaviour etc. Their research identified tweet attributes that can be used to detect accounts that exhibit abnormal behaviour (e.g. posting spam or malicious URLs). Cao and Caverlee analyzed the behaviour of Twitter users to detect tweets classified as spam, using meta-data from the user account posting the spam or URL and the user account clicking the URL \citep{Cao2015}. Their hypothesis was based on the assumption that it is difficult to manipulate such behavioural signals.  
 Chen et al. \citep{chen2016} used a Finite State Machine based spam template, demonstrating that a cyber criminal can create 2000 tweets from a single template and discovering that such users were using multiple accounts to post spam in a coordinated manner to avoid detection. They were exhibiting "load balancing" - a  technique frequently used to prevent denial of service attack-  but in this case posting from multiple account to prevent being detected. 
Stringhini et. al. created honey profiles on the top three OSNs and recorded the content and interactions made to these profiles to identify tweet attributes contributing to malware propagation \citep{Stringhini2010}. 
Benevenuto et al. focused on identifying spam centred around Twitter exclusively by using twenty three tweet attributes \citep{Benevenuto2010}. Grier et al. analysed spam behaviour and the effectiveness of using a blacklist of URLs to detect spam on Twitter\citep{Grier2010}. They analysed spam behaviour of tweets by  recording the frequency of tweet being posted. 
Yang et al. \citep{Yang2009} used features based on timing and automation to detect spam on Twitter. Their research was focused on the relationship such as betweeness centrality and bidirectional link ratio between spam nodes and their neighbouring nodes. The same authors collaborated with Zhang and Shin\citep{chaoyangrobertharkreaderjialongzhangseungwonshinguofeigu2012} to analyse the cyber criminal ecosystem on Twitter studying inner and outer social relationships. The inner social relationship hypothesised that criminal accounts are interconnected. The outer social relationship highlighted those accounts that follow a criminal account and help each criminal account to be well hidden in the network. Similarly a feature based approach was employed by \citep{Cresci2015} by building a classifier to detect fake accounts created by cyber criminals to inflate the number of followers. 
\\
To date the research has been focused on studying OSN accounts and URL characteristics to identify those tweets or accounts that are exhibiting deviant behaviour (posting spam or malicious URLs). Providing evidence that OSN accounts or URLs may be malicious can be beneficial but given the frequency and volume at which new accounts emerge, the only way to determine actual malicious behaviour is occurring is to observe it. Once malicious activity occurs it is currently not possible to flag it and stop it. None of the methods published to date allow us to observe malicious activity and block it to minimise the damage. Thus we propose to build on the existing literature that uses characteristics as features and include them in a predictive model that will incorporate tweet attributes to predict that the URL is likely to perform malicious activity during the early stages of interaction, providing a novel enhancement to the research field whereby we can observe malicious behaviour, including that of newly created accounts with limited account history, and block it before maximum damage occurs.  
\subsection{Detecting malicious content by analysing the static or dynamic activity of a Web page}There are two ways to analyse the activity of a Web page. Static analysis looks at the code that drives the page, looking for recognised malicious code and methods. Dynamic analysis executes the code by interacting with the Web page and observes the behaviour on the endpoint and on the local system, also looking for evidence of known malicious activity but also enabling a more broad analysis of observable activity. 
\textit{Static analysis: }McGrath and Gupta analysed the anatomy of phishing URLs, studying the patterns of characters and domain length in URLs to develop a filter to detect phishing URLs \citep{McGrath2008}.
In a similar approach an automated classification model was built based on lexical and host based features to detect malicious URL using statistical models \citep{Ma2009a}.
Canali et. al. developed a filter called \textit{Prophiler} \citep{Canali2011} that uses features derived from URLs and Web page code to determine whether a drive by download will occur. In another approach Kapravelos et al. compared similarities between various JavaScript programs to detect malicious Web pages \citep{Kapravelos2013}. \\
\textit{Dynamic analysis: }A system was develop by \textit{Cova et al}  to detect malicious Web pages in two stages\citep{Cova2010}.  In the first stage various features such as URL redirects, length of dynamic code, number of dynamic execution class etc. were used to detect an anomaly. 
In the second part they used a custom built browser to open the URL and record the events used to detect malicious behaviour. Building on the principle of detecting malware by analysing dynamic execution of code Kim et al. proposed a model to systematically explore possible execution paths and reveal malicious behaviours based on the execution paths \citep{kim2017j}. This is achieved by analysing function parameters that could expose suspicious DOM injection and reveal malicious behaviour. In a similar approach, Javasinghe et al. used the dynamic behaviour of a Web page to detect a drive by download attack \citep{jayasinghe2014efficient}. Abobe Flash animations are a well known entry point for Web-based attacks and these have been studied at various levels during the interpreter’s loading and execution process to detect malicious code \citep{wressnegger2016comprehensive}.
Research has also been undertaken to build a machine classifier based on network activity to detect malware. In one approach Bartos and Sofka looked at network traffic to build the classifier from data captured in the form of proxy logs generated by 80 international companies \citep{bartos2016optimized}. By doing so they were able to detect both known as well as previously unseen security threats based on network traffic. Similarly, Burnap et al. built a real time classifier specific to drive by downloads originating from Twitter based on network activity and machine activity \citep{Burnap2014}. Looking at the dynamic redirection of Web pages has been proposed to detect phishing and spamming webpages in \citep{lee2013warningbird, Lee2011}.
This was extended to using forward and graph based features in \citep{Cao2016}.\\
In summary, while excellent results have been achieved by studying the static or dynamic activity of a Web page, the focus has been on \textit{detection}. As stated at the end of the previous section, to identify malicious activity in OSN it must be observed, and generally once it is observed, it is a problem that needs to be remedied. As with the research in the previous section, none of the research to date that focuses on Web page activity has proposed a model capable of observing and potentially blocking malicious activity. Thus, in this paper we focus on \textit{prediction}, proposing a model that can classify a URL into malicious or benign based on OSN account attributes (as per the previous section) and also dynamic machine behaviour - activity observed when the URL is clicked and the Web page is being loaded. The aim is to predict that behaviour observed in the early stages of loading a Web page is likely to lead to malicious activity at a later stage - providing new capability for a user to block the completion of the malicious actions rather than depend on detection and repair at a significant cost and inconvenience.

\section{Experimental Setup}

\subsection{Data Collection}
We collected data around two popular sporting events that were expected to attract a large number of users. This made them potential targets for cyber criminals to carry out drive by download attacks. For our experiments we identified the European Football Championships (\#Euro2016) and the Olympics (\#Rio2016) in 2016. Both generated some of the largest volumes of tweets in 2016. Tweets containing a URL and hashtags relating to these events were captured via the Twitter streaming API. The rationale behind selecting two events was to the determine whether our predictive model would generalise beyond a single event and be applicable for use on URLs posted around other events.
For Euro 2016 we  captured tweets from the period of 10 June to 14 July 2016 using the hashtag \#Euro2016. We harvested 3,154,605 tweets that contained a URL. During the opening ceremony that marked the opening of the Olympics in 2016 (the peak of public interest) we captured 148,881 tweets that contained a URL using the the hashtag \#Rio2016. From the captured tweets we selected a systematic sample of 7500 unique tweets to identify 975 malicious URLs for European Football Championships dataset and around 5000 tweets were used to identified around 525 malicious unique tweets for Olympics 2016 dataset by using a high interaction client side honeypot. High interaction honeypots perform dynamic analysis of interaction behaviour between a client machine and that of a Web server. For our experimental results we used CaptureHPC toolkit \cite{CaptureH33:online}. CaptureHPC operates by visiting each URL that is passed to it through a virtualised Sandboxed environment - interacting with the Web page for a pre-defined amount of time. At the end of the interaction time Capture HPC determines if any system-level operations have occurred including file, process and registry changes made to the system. Based on these changes it classifies that URL into malicious or benign \citep{puttaroo2014challenges}. The classification is based on an \textit{exclusion list} that is created based on known file, process or registry entries that are targeted by drive by download attacks. This exclusion list is updated every 14 days to reflect the most recent actions that have been observed in drive by download attacks. The exclusion list contain rules that are identified while visiting malicious or benign Web sites. CaptureHPC therefore gives us a label we can use for supervised learning and a set of activity logs we can use to train a system to recognise the 'early warning signals' that are present \textit{before} the exclusion list flag would have been raised. 

 \begin{figure*}[htb]
            
            \centering
            \includegraphics[width=\textwidth]{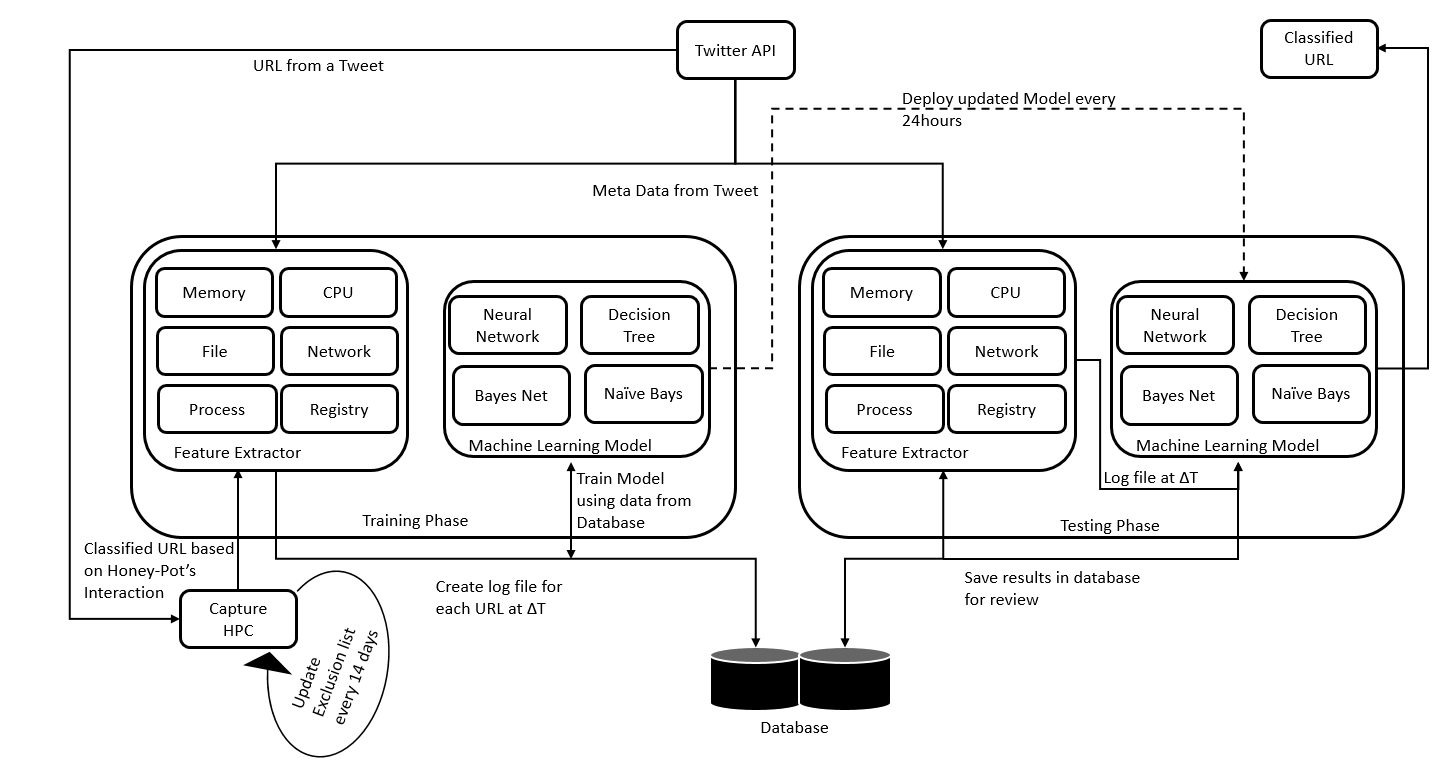}

            \caption{Architecture of Predictive Model}\label{fig:archi}
\end{figure*}

\subsection{Architecture of the Predictive Model}

The predictive model has three main components (see Fig. \ref{fig:archi}): \textit{feature extraction}, \textit{persistent storage} and \textit{machine learning}. The main function of feature extraction is to create a timeline of measurable observations on the client system in terms of machine activity and tweet attributes from the time a URL is opened to the point at which a drive by download is carried out or system becomes idle. The feature extractor opens each URL that is passed to it in a Sandboxed environment and starts creating snapshots of machine activity at time interval \textit{'t'} for a period of \textit{'p'}. For our experiment, \textit{t=1 second} and the observation period is defined as \textit{p=10 seconds}. The first snapshot is generated when a URL is 'clicked' at \textit{t=1 second}, and then subsequently at an interval of $t$. Each snapshot is written to a database for persistence as the Sandboxed environment is wiped clean after each URL has been visited. Each database insert includes (i) machine activity and (ii) meta data of the tweet containing the URL. For machine activity we log 54 metrics including network activity, file, process, registry, RAM use, CPU usage (see Appendix 1 for a longer list and associated Pearson correlation scores with the malicious/benign class). We also use 24 pieces of meta data from the tweet, including user name, user screen name, user id, follower count, friends count, and age of account (see Appendix 1 for longer list). This produces 78 attributes every second for a period of $p$. During the training phase we know whether the URL is malicious or benign based on the results from CaptureHPC. This label is inserted into the database with each snapshot. 
Once the observation time is complete, the Sandboxed environment is reset to a malware-free state so that each new URL can be opened in a known malware free configuration with a consistent baseline. 
\\
The third component is the machine learning phase. For our predictive model we trained four different machine algorithms to determine the best method for class prediction using these data. We used the Weka toolkit to compare the predictive accuracy of (i) generative models that consider conditional dependencies in the dataset (BayesNet) or assume conditional independence (Naive Bayes), and (ii) discriminative models that aim to maximise information gain (J48 Decision Tree) and build multiple models to map input to output via a number of connected nodes, even if the feature space is hard to linearly separate (Multi-layer Perceptron). To test the models we used the feature extractor and the learned machine learning model from the training phase. Tweets from the testing dataset (in the first instance using 10-fold cross validation, and later using a holdout testing dataset) were passed into the feature extractor, which opened the URL in the Sandboxed environment and created the machine activity and tweet meta-data snapshots at every time interval. Each snapshot was passed onto the learned model which classified the snapshot as malicious or benign. If the result was 'benign', the process continues to the next snapshot. The first time the outcome is 'malicious', the process stops and the URL is classified as malicious, killing the connection to the Web page. \\
The framework is designed to be adaptive to an ever-changing environment by periodically updated the labelling method used to train and test the classifier so that new malware behaviour is reflected in the labels. This is achieved by periodically updating the exclusion list of the honeypot. The exclusion list is updated once every 14 days by running URLs in CaptureHPC after executing them in known malware labelling Web sites like \textit{Virustotal} \citep{VirusTot92:online}, which provide labels based on the leading commercial anti-virus tools. Based on the machine activity observed in terms of files/process/registry we update the exclusion list \citep{puttaroo2014challenges}.

\section{Results}
 \begin{figure*}[!htb]
            \centering
            \begin{subfigure}[t]{0.40\textwidth}
                \includegraphics[width=\textwidth]{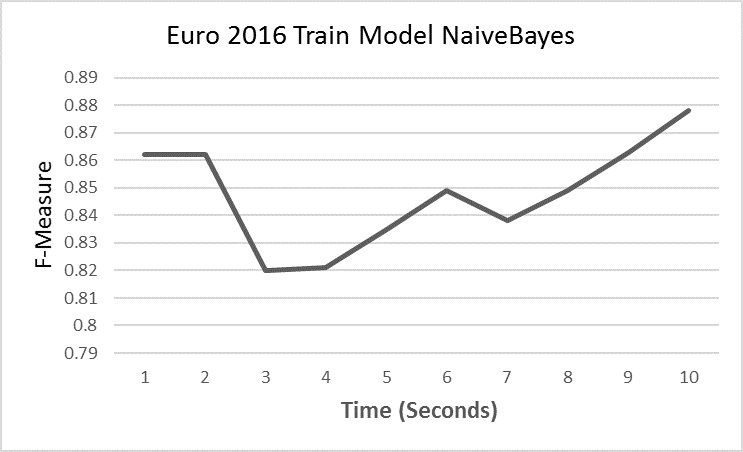}
                \caption{F-Measure of Naive Bayes over time during training phase}
                \label{fig:train_naive}
            \end{subfigure}
            \qquad 
            \begin{subfigure}[t]{0.40\textwidth}
                \includegraphics[width=\textwidth]{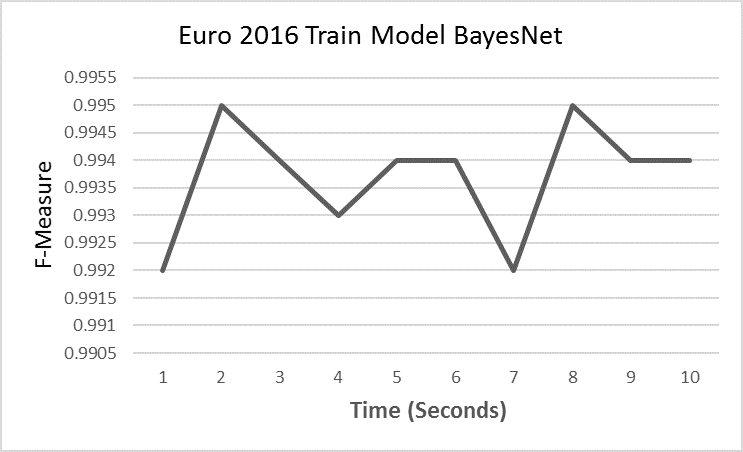}
                \caption{F-Measure of BayesNet over time during training phase}
                \label{fig:train_bayes}
            \end{subfigure}
    
            \begin{subfigure}[t]{0.40\textwidth}
                \includegraphics[width=\textwidth]{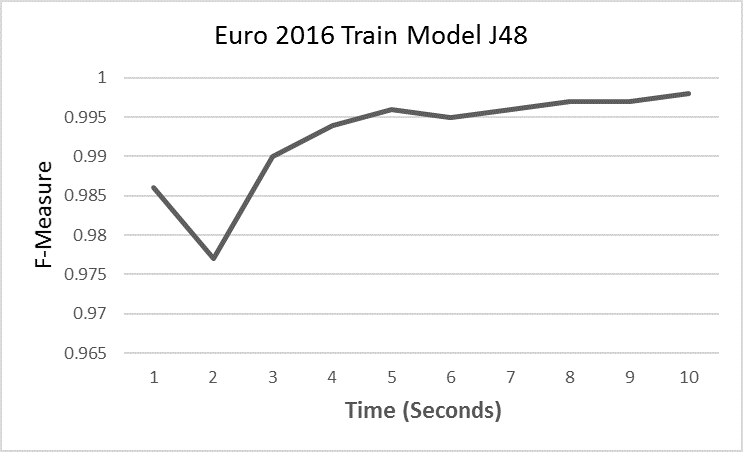}
                \caption{F-Measure of J48 over time during training phase}
                \label{fig:train_j48}
            \end{subfigure}
            \qquad
            \begin{subfigure}[t]{0.40\textwidth}
                \includegraphics[width=\textwidth]{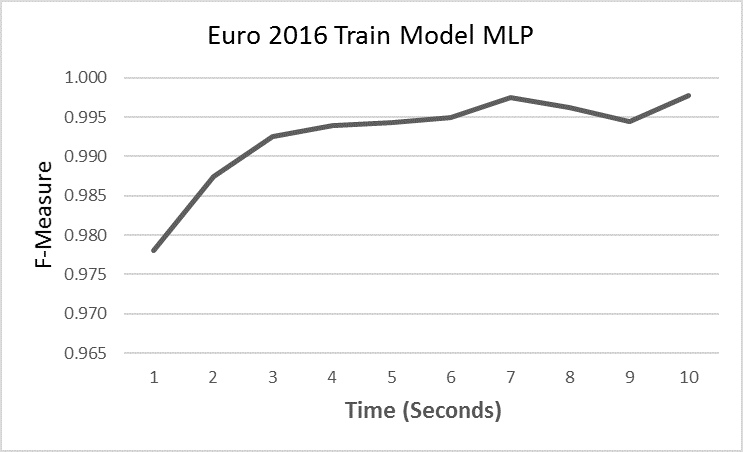}
                \caption{F-Measure of MLP over time during training phase}
                \label{fig:train_mlp}
            \end{subfigure}
            \caption{F-Measure of all machine learning algorithm over time during training phase}\label{fig:results}
            
\end{figure*}

\begin{figure}[ht]
            \centering
            \includegraphics[width=0.60\textwidth]{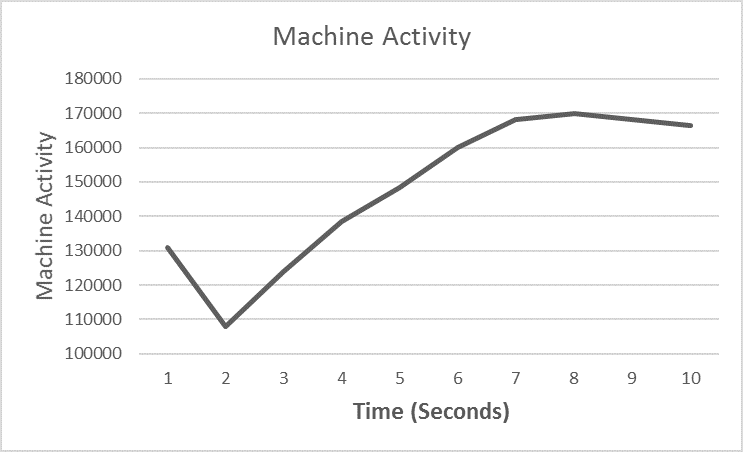}
            \caption{Machine activity over time}
            \label{fig:machineact}
\end{figure}
\subsection{Training on data from Euro 2016}
To determine which models provide the best predictive power - not just overall classification accuracy on all data - each model was trained and tested using data from sequential, cumulative time intervals. That is, at each time interval $t$ from $t=1$ to $t=p$ where $p$ is the total number of time intervals (in this case $p=10$), each model was trained and tested using data from \textit{t=1-to-p} where \textit{p=p+1}. Each interval was evaluated with ten fold cross validation using the Weka toolkit. The results were calculated using standard classification metrics - Precision, Recall, and F-Measure.\\
The results for for each classifier are presented in  Figure \ref{fig:results}. In each sub-figure, the machine learning model is trained and tested on the metrics derived using the Euro2016 data-set. Time in each table represents the time in seconds elapsed from the time the URL was clicked, and the starting point is defined as $t=1$. For example \textit{Time=2} means 1 second has elapsed since the URL has been 'clicked' (URL clicked at $t=1$). 
Models built using the Naive Bayes and J48 algorithms (see Figures \ref{fig:train_naive} and  \ref{fig:train_j48}) exhibit similar behaviour - they both have a dip in accuracy from the starting point and then it gradually continues to rise up. One explanation for this could be that during early seconds there is a lack of system activity (see Figure \ref{fig:machineact}), leaving the algorithm  struggling to differentiate between benign and malicious activity. The F-measure of the J48 machine learning model follows the trend of machine activity and continues to rise as more activity is recorded. When we compare the generative probabilistic models (Naive Bayes and BayesNet) we find that BayesNet outperforms NaiveBayes, suggesting interdependencies between attributes. This is logical as, for instance, when malicious network activity occurs is likely that CPU and RAM use will also spike due to additional resource being required for the activity. Looking at the results of the MLP model (see Figure \ref{fig:train_mlp}) we see the model is able to better weight the machine activity and tweet meta-data to control for the lack of machine activity at the start of the interaction. The F-measure rises smoothly from 1 second, suggesting it is making better use of the Twitter metadata to improve accuracy in the early stages of activity.
In terms highest F-measure achieved, the J48 and MLP models perform best with 0.998 at 10 seconds. At 3 seconds the results are almost identical. The key difference between models being a slight improvement in MLP at 2 seconds, but this is countered by the speed at which J48 returns a result. The MLP result takes longer than a second to be returned, whereas the J48 takes milliseconds. Thus, in practical application, the J48 model is most likely to be favourable. 


\subsection{Training Model without Online Social Network Platform attributes} 
\begin{figure}[htb]

            \centering
                \includegraphics[width=0.60\textwidth]{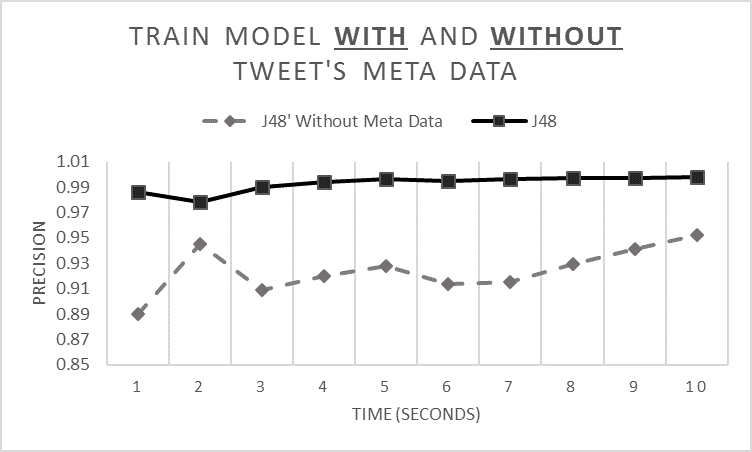}
                \caption{Train J48 Model without OSN meta data}
                \label{fig:meta}
\end{figure}

A lot of research has been done in the past to detect malicious/spam tweets propagating on Twitter based on tweet attributes \citep{Stringhini2010, Benevenuto2010, Grier2010, chaoyangrobertharkreaderjialongzhangseungwonshinguofeigu2012, Yang2009, Cresci2015}. Thus we included tweet meta data as part of the feature set for prediction in the previous section. However, these features are quite ideosyncratic and not consistent across different OSNs. For instance, if we wanted to predict a drive by download via other OSNs such as Facebook, Tumblr or Instagram, we would get a slightly different set of user characteristics from the metadata available. Thus, we aimed to determine the impact of removing these features and use machine activity data alone to determine the applicability of our method across different OSNs. To conduct this experiment we selected the model from the previous experiment that provided us the best performance - the J48 algorithm that displayed apparent correlation with machine activity. We  retrained the model using only the machine activity - no tweet metadata. Table \ref{j48machine} and Figure \ref{fig:meta} show performance of the of the model over time.

\begin{table}[htb]
\centering
\caption{Training Model On Euro 2016 log file using J48 Algorithm without Tweet Meta data}
\label{j48machine}
\begin{tabular}{|l|l|l|l|}
\hline
\multicolumn{4}{|c|}{\textbf{\begin{tabular}[c]{@{}c@{}}Euro 2016 Train Model -J48 \\ (Without Tweets Meta data)\end{tabular}}} \\ \hline
\textbf{Time}               & \textbf{Precision}               & \textbf{Recall}              & \textbf{F-Measure}              \\ \hline
1                           & 0.89                             & 0.863                        & 0.858                           \\ \hline
2                           & 0.945                            & 0.94                         & 0.939                           \\ \hline
3                           & 0.909                            & 0.9                          & 0.901                           \\ \hline
4                           & 0.92                             & 0.904                        & 0.905                           \\ \hline
5                           & 0.928                            & 0.916                        & 0.915                           \\ \hline
6                           & 0.914                            & 0.899                        & 0.897                           \\ \hline
7                           & 0.915                            & 0.899                        & 0.897                           \\ \hline
8                           & 0.929                            & 0.918                        & 0.918                           \\ \hline
9                           & 0.941                            & 0.933                        & 0.933                           \\ \hline
10                          & 0.952                            & 0.947                        & 0.947                           \\ \hline
\end{tabular}
\end{table}

Figure \ref{fig:meta} shows the precision metrics for the J48 model when trained with and without tweet meta data. When we compare the precision of both J48 models we observe that the model built solely on machine activity data fluctuates over time. The model F-measure drops by around 13\% at t=1 second. This suggests that Twitter's idiosyncratic attributes such as number of followers significantly contribute to accurate classification of malicious URLs but that the model is still highly accurate when using machine activity alone, making it likely that the approach would work to detect drive-by-downloads on other OSNs. 

Without the OSN metadata the model seems able to better cope with the low rate of activity at the start of the interaction, which is interesting as this is the opposite of the situation when metadata were used to train the model. The key finding here is that including the OSN metadata improves the prediction of the classifier by 12.98\%, thus in future our aim will be to try and retain user account characteristics where possible when applied to OSNs outside of Twitter - but that our model still provide a high predictive performance even without these data, providing promising results for the application of machine activity models for predicting malicious behaviour in URLs on multiple OSN platforms. 
\begin{table}[htb]
\centering
\caption{Test Model On Olympics 2016 Dataset}
\label{vote}
\begin{tabular}{|l|l|l|l|}
\hline
\multicolumn{4}{|c|}{\textbf{\begin{tabular}[c]{@{}c@{}}Test Olympics -Vote Algorithm \\ (NaiveBayes and J48)\end{tabular}}} \\ \hline
\textbf{Time}              & \textbf{Precision}              & \textbf{Recall}              & \textbf{F-Measure}             \\ \hline
1                          & 0.84                            & 0.733                        & 0.74                           \\ \hline
2                          & 0.871                           & 0.841                        & 0.839                          \\ \hline
3                          & 0.879                           & 0.854                        & 0.852                          \\ \hline
4                          & 0.873                           & 0.845                        & 0.842                          \\ \hline
5                          & 0.881                           & 0.859                        & 0.856                          \\ \hline
6                          & 0.882                           & 0.865                        & 0.862                          \\ \hline
7                          & 0.847                           & 0.83                         & 0.825                          \\ \hline
8                          & 0.806                           & 0.801                        & 0.798                          \\ \hline
9                          & 0.724                           & 0.722                        & 0.722                          \\ \hline
10                         & 0.637                           & 0.619                        & 0.616                          \\ \hline
\end{tabular}
\end{table}
\begin{figure}[t]
            \centering
                \includegraphics[width=0.60\textwidth]{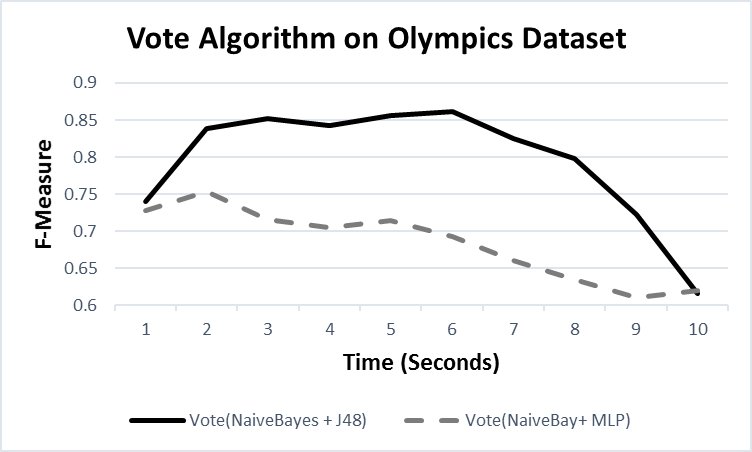}
                \caption{Testing on Olympics Data using model built earlier}
                \label{fig:test}
\end{figure}

\subsection{Testing using unseen data from Olympics 2016 }
In the previous two experiments we validated our predictive models using a single dataset from Euro 2016 and obtained promising results. One possible limitation with this experiment is that cyber attack methods vary over time. For instance, in a second unrelated event we may see a new collection of individuals spreading malicious URLs, and indeed a different behavioural profile exhibited by the URLs. We therefore now introduce an unseen dataset from the Olympics 2016. This dataset has played no part in training the model so is completely unseen, testing the generality of the approach to some degree. 
Given that J48, MLP and Naive Bayes (NB) models performed best on the Euro 2016 data, we combined these using a Vote meta-classifier. The \textit{Vote} algorithm allows two or more machine learning algorithms to be combined in such a way that the the label likelihood from each model is used to provide the classification label for each test instance. In our case we used the average probability as the decision point. Through experimentation we narrowed down two combinations of methods that produced the highest accuracy: J48 \& NaiveBayes and NaiveBayes \& MLP. Figure \ref{fig:test} shows the F-measure for both. The combination of J48 with NaiveBayes reaches an F-measure of 0.85 after just two seconds into the interaction with a Web page. Note again that t=1 is the time the test machine launches the URL so there is a lag of 1 second, meaning t=3 is actually 2 seconds after the URL is clicked. The NaiveBayes and MLP combination reaches a maximum F-measure of 75.3\%. Thus there is a significant performance difference when combining the Naive Bayes and J48 models. This is somewhat counter intuitive given the MLP and J48 algorithms were almost indistinguishable at 3 seconds in the previous experiments, and that J48 is a rule-based model. We would expect a rule-based model to overfit to a single event (i.e. the CPU, RAM and network traffic would have a large variance between events as demonstrated by \cite{Burnap2014}). This was not the case, and in fact this combination produced a model that is capable of detecting malicious URLs in an unseen dataset with 85\% accuracy at only 2 seconds into the interaction. 

\begin{figure}[t]
            \centering
            \includegraphics[width=0.60\textwidth]{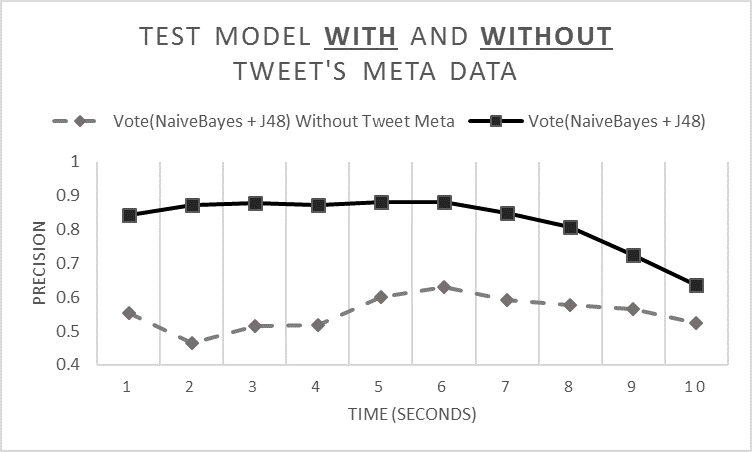}
            \caption{Comparison of Results on Unseen Data with and without tweet metadata}
            \label{fig:test_withandwithout}
\end{figure}

We next rebuilt the Vote model with and without tweet meta data meta data. Figure \ref{fig:test_withandwithout} shows the result of the classifier when we tested this model on the Olympics 2016 (unseen) dataset. We see a significant increase (on average an increase of 32\% was observed) in precision of the classifier when tweet attributes were added to machine data. This suggests that even though there is similarity in tweet attributes across events they are not enough to accurately classify a URL on their own, and we still require machine data to improve our classification across events. Note also that the the results of the same models based on tweet meta data alone using the Olympics 2016 dataset gave an F-measure of only 16\% (full results not shown for brevity). We can see that while the attack vectors as measured by system activity are changing between events (hence the drop in performance when remove the Twitter metadata), the combination of network characteristics of the individuals posting malicious URLs, and machine activity recorded while interaction with URLs, remain fairly stable - showing a drop in F-measure from 0.977 to 0.839 at 2 seconds between events. Our model may therefore not be limited to a single case, but could be applied to multiple events on Twitter maintaining reasonably low error rates when predicting malicious URLs just 2 seconds into the interaction.


\begin{figure}[t]
            \centering
            \includegraphics[width=0.60\textwidth]{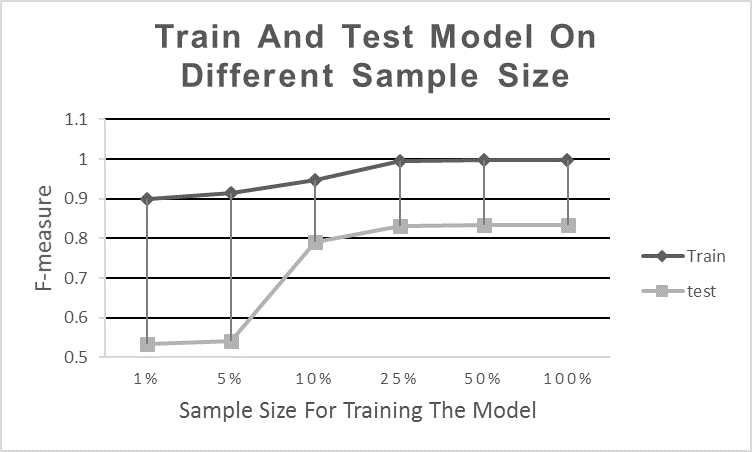}
            \caption{Comparing classifier accuracy in terms of F-measure when data set is changed}
            \label{fig:machine}
\end{figure}
\subsection{Adaptive nature of the predictive model}
To make our predictive model adaptive, a feed-forward architecture was implemented (see Figure \ref{fig:archi}). The rationale was to ensure that new techniques employed by cyber criminals to carry out a drive by download attack, as captured in the form of machine activity, are continually captured and considered while training the model. In order to check the effectiveness of the feed-forward architecture in achieving this we conducted a further experiment. We trained the model on the Euro 2016 dataset with varying sample sizes, and tested using 10 fold cross validation. We then tested the model on an unseen dataset (Olympics 2016), with the hypothesis that increasing the size of a dataset would capture new machine behaviour that would increase the diversity of features seen by the model and improve the overall F-measure of the predictive model.\\  
We used a range of sample sizes for model training - 1\%,5\%, 10\% ,25\%, 50\% and 100\%. Figure \ref{fig:machine} displays the results of these experiments.

We found that while training the model with only 1\% of total sample size, using 10 cross fold technique, it produced an F-Measure of 0.89. However, when we tested the model on an unseen dataset we found the F-measure dropped to 0.533. By increasing the size of the training dataset from 1\% to 100\% in various stages  we aimed to simulate how the model would behave as new data is added to the model time and the feature diversity increases. We observed that the F-measure did indeed increase with increases in dataset size during the training phase as well as with the testing phase, showing the model to be adaptive when observing more diverse machine behaviour. We saw a significant jump in the F-measure (from 0.54 to 0.80) when the sample size was increased to 10\%. 
However, little change in the F-measure was observed when we increased the sample size from 25\% to 100\%, suggesting that 25\% of data representing machine activity is enough to build a model that will give us over 83\% F-measure. After this point more data does not appear to improve prediction accuracy. 

\section{Conclusions}
As Online Social Networks (OSNs) become a key source of information publication and propagation following global events it has become an environment that is particularly vulnerable to cyber attack via the injection of shortened URLs that take the user to a malicious server from which a 'drive by download' attack on the local machine is launched. In this paper we aimed to build on a body of work that has developed methods to identify malicious URLs in OSNs in an effort to combat the problem. Existing work has developed methods to provide evidence that OSN accounts or URLs may be malicious, which can be beneficial, but given the frequency and volume at which new accounts emerge, the only way to determine actual malicious behaviour is occurring is to observe it. Once malicious activity occurs it was previously not possible to flag it and stop it. None of the methods published prior to this work allowed us to observe malicious activity and block it to minimise the damage. The main focus of our research was therefore to develop a method capable of identifying a URL as malicious or benign based on machine activity metrics generated and logged during interaction with a URL endpoint, and OSN user account attributes (in this case Twitter users) associated with the URL. Furthermore, the aim was to \textit{predict} that the URL was likely to be malicious within seconds of opening the interaction - before the drive by download attack could complete the execution of its payload. This is the first time a method has been tested to predict a malicious outcome before it actually takes place - existing literature always classified URLs using all the data generated throughout an interaction period - so provided a post-hoc result, or without actually observing the malicious activity - making a decision based on previously observed behaviour . 

We captured tweets containing URLs around two global sporting events. Our system produced a second-by-second time series of system-level activity (e.g. CPU use, RAM use, network traffic etc.) during visitation of a Web page. We trained the classification model using four different types of machine learning algorithm on log files generated from one event (Euro 2016). The model was then validated using tweets captured during another event (Olympics 2016). The rationale was to determine if similar machine activity and tweet attributes were exhibited in two completely different events (i.e. does the model generalise beyond a single event). A ten fold cross validation was performed to train the model and an accuracy of around 99\% was achieved by using the log files generated at 1 second into the interaction with a Web server, and a maximum of 99.8\% at time=10 seconds. One of the interesting observations during training phase was that by using tweet attributes we can increase the accuracy by 12.98\% during training and around 32\% during testing phase when compared to machine activity alone, showing that the Twitter metadata exhibited by cyber criminals to carry out drive by download attacks were relatively stable, while the URL behaviour changed.\\
When tested using an unseen dataset (Olympics 2016) we achieved an accuracy of 83.9\% from log files generated at 2 seconds - that is 1 second after launching the URL. The highest accuracy achieved on the unseen event was 86\% at around 4 seconds from the time the URL was launched. Our model may therefore not be limited to a single case, but could be applied to multiple events on Twitter maintaining reasonably low error rates when predicting malicious URLs just 1 second into the interaction. The model allows us to reduce the detection time of a malicious URL from minutes - the time taken to run the URL in a secure sandbox environment - to 4 seconds, with an accuracy of 86\% on an unseen dataset. Furthermore it allows us to stop the execution process with 84\% accuracy just 1 second after clicking the URL, preventing full execution of the malicious payload, rather than detecting the malicious action retrospectively and having to repair the system.
Future work includes increasing the granularity further by creating log files at shorter intervals to determine if we can detect malicious URLs even earlier in the execution cycle, to avoid the key limitation which is that a cyber criminal can evade detection if the connection is dropped within one second.

\section{References}
\bibliographystyle{elsarticle-num}
\bibliography{bib}
\section{Appendix 1}

\begin{table}[htb]
\centering
\caption{Feature Selection of Attributes using Pearson's R Correlation between attributes and its class (Malicious/benign) }
\label{my-label}
\resizebox{\columnwidth}{!}{
\begin{tabular}{|l|l|l|l|l|l|l}
\cline{1-6}
\textbf{Sr No} & \textbf{Pearsons Correlation} & \textbf{Attribute Name} & \textbf{Sr No} & \textbf{Pearsons Correlation} & \textbf{Attribute Name} &  \\ \cline{1-6}
1 & 0.4059 & Process Create Time & 32 & 0.0184 & retweet user screem name &  \\ \cline{1-6}
2 & 0.2950 & disk io counter write bytes & 33 & 0.0175 & user time zone &  \\ \cline{1-6}
3 & 0.2915 & disk memory free & 34 & 0.0172 & retweet user verified &  \\ \cline{1-6}
4 & 0.2915 & disk memory used & 35 & 0.0151 & disk io counter read times &  \\ \cline{1-6}
5 & 0.2914 & disk memory percent & 36 & 0.0138 & user language &  \\ \cline{1-6}
6 & 0.2621 & CPU & 37 & 0.0137 & process id net &  \\ \cline{1-6}
7 & 0.1125 & user verified & 38 & 0.0133 & age &  \\ \cline{1-6}
8 & 0.0981 & virtual memory percent & 39 & 0.0132 & retweet user timezone &  \\ \cline{1-6}
9 & 0.0974 & virtual memory available & 40 & 0.0103 & retweet favourite tweet count &  \\ \cline{1-6}
10 & 0.0974 & virtual memory free & 41 & 0.0091 & retweet user id &  \\ \cline{1-6}
11 & 0.0974 & virtual memory used & 42 & 0.0082 & disk io counter write times &  \\ \cline{1-6}
12 & 0.0939 & Packets received & 43 & 0.0074 & process username &  \\ \cline{1-6}
13 & 0.0935 & Bytes received & 44 & 0.0072 & retweet user favourites count &  \\ \cline{1-6}
14 & 0.0891 & disk io counter read bytes & 45 & 0.0069 & memory percent &  \\ \cline{1-6}
15 & 0.0885 & swap memory free & 46 & 0.0063 & process path &  \\ \cline{1-6}
16 & 0.0885 & swap memory used & 47 & 0.0062 & process name &  \\ \cline{1-6}
17 & 0.0874 & swap memory percentage & 48 & 0.0061 & process status &  \\ \cline{1-6}
18 & 0.0799 & Packets Sent & 49 & 0.0061 & remote ip &  \\ \cline{1-6}
19 & 0.0647 & disk io counter write count & 50 & 0.0061 & connection Establish listen &  \\ \cline{1-6}
20 & 0.0638 & user friends count & 51 & 0.0061 & user coordinates &  \\ \cline{1-6}
21 & 0.0627 & disk io counter read count & 52 & 0.0047 & process id &  \\ \cline{1-6}
22 & 0.0617 & Bytes Sent & 53 & 0.0043 & source path &  \\ \cline{1-6}
23 & 0.0548 & user name & 54 & 0.0036 & cmd line statement &  \\ \cline{1-6}
24 & 0.0495 & retweet user name & 55 & 0.0036 & process exe path &  \\ \cline{1-6}
25 & 0.0368 & retweet retweet count & 56 & 0.0036 & cpu time user &  \\ \cline{1-6}
26 & 0.0292 & user screen name & 57 & 0.0034 & retweet user friends count &  \\ \cline{1-6}
27 & 0.0261 & user location & 58 & 0.0023 & retweet user followers count &  \\ \cline{1-6}
28 & 0.0234 & user followers count & 59 & 0.0010 & cpu time system &  \\ \cline{1-6}
29 & 0.0214 & type & 60 & 0.0006 & port number &  \\ \cline{1-6}
31 & 0.0186 & retweet user location & 61 & 0.0000 & swap memory swap in &  \\ \cline{1-6}
\end{tabular}}
\end{table}

\end{document}